\documentclass{ws-mpla}
\usepackage[super]{cite}
\usepackage{graphicx}
\usepackage{wrapfig}% wrap text around images
\usepackage{float}
\usepackage{multirow}
\usepackage[group-separator={,},group-minimum-digits=4]{siunitx}
\usepackage{comment}
\usepackage{booktabs,tabularx}
\usepackage{ amssymb }
\usepackage{gensymb}
\usepackage[utf8]{inputenc}
\usepackage{xcolor}
\usepackage[T1]{fontenc}

\DeclareSIUnit\GeV{\giga\eV}
\DeclareSIUnit\MeV{\mega\eV}
\DeclareSIUnit\keV{\kilo\eV}
\DeclareSIUnit\eVperc{\eV\per\clight}
\DeclareSIUnit\clight{\text{\ensuremath{c}}}
\DeclareSIUnit\MeVperc{\mega\eV/\clight}
\DeclareSIUnit\keVperc{\kilo\eV/\clight}

\DeclareSIUnit\MHz{\mega\Hz}
\DeclareSIUnit\us{\micro\second}
\DeclareSIUnit\ns{\nano\second}

\begin{document}

\markboth{Manolis Kargiantoulakis}
{A Search for Charged Lepton Flavor Violation in the Mu2e Experiment}

\newcommand{\etal}{\textit{et al}. }
\newcommand{\ie}{\textit{i}.\textit{e}. }
\newcommand{\eg}{\textit{e}.\textit{g}. }
\newcommand{\Lagr}{\mathcal{L}}

%\graphicspath{{./fig/}}

%%%%%%%%%%%%%%%%%%%%% Publisher's Area please ignore %%%%%%%%%%%%%%
%%%%%%%%%%%%%%%%%%%%%%%%%%%%%%%%%%%%%%%%%%%%%%%%%%%%%%%%%%%%%%%%%%%

\title{\textbf{A Search for Charged Lepton Flavor Violation in the Mu2e Experiment}}

\author{\footnotesize Manolis Kargiantoulakis~\footnote{ekargian@fnal.gov}
}

\address{Fermi National Accelerator Laboratory\\
Batavia, Illinois 60510, United States of America
\\
ekargian@fnal.gov}

\maketitle

%\pub{Received (Day Month Year)}{Revised (Day Month Year)}

\begin{abstract}
The Mu2e experiment will search for the neutrino-less conversion of a muon into an electron in the field of an aluminum nucleus. 
An observation would be the first signal of charged lepton flavor violation and de facto evidence for new physics beyond the Standard Model.
The clean signature of the conversion process offers an opportunity for a powerful search: Mu2e will probe four orders of magnitude beyond current limits, with real discovery potential over a wide range of well motivated new physics models.
This goal requires an integrated system of solenoids that will create the most intense muon beam in the world, and suppression of all possible background sources.
The Mu2e components are currently being constructed, with the experiment planned to begin operations in the Fermilab Muon Campus within the next few years.

\end{abstract}

\ccode{PACS Nos.: 11.30.Hv; 12.15.Ff; 13.35.Bv; 14.60.Ef}

\section{Introduction and Context}	

For all its successes over many decades, the Standard Model (SM) of physics is viewed only as an effective low energy approximation and we are motivated to search for a more fundamental underlying structure.
Through precision searches we are granted indirect access to higher energy scales or more weakly coupled interactions, and signals of physics beyond the Standard Model (BSM) may have already been detected. 
The observation of neutrino oscillations constitutes a de facto extension to the standard theory, with remaining questions regarding the nature of neutrinos;
the muon anomalous magnetic moment measurement has yielded a $>3\sigma$ discrepancy from the SM prediction,\cite{Bennett_2006}
an important result that has guided experimental searches and BSM theory;
and numerous signals in semi-leptonic $B$ meson decays offer corroborating hints for violation of lepton flavor universality over a multitude of channels,\cite{BABAR_2012,Belle_2014,LHCb_2015,Belle_2016,Belle_2017,LHCb_2018}
with combined statistical significance around the $4 \sigma$ level.\cite{Graverini_2019}
This pattern of experimental deviations reveals precision searches in the flavor sector as a promising and well motivated field.

\subsection{Charged lepton flavor violation}

The concept of flavor in particle physics begins with the discovery of the muon by Anderson and Neddermeyer in cosmic radiation\cite{Neddermeyer_1937} and the realization that it appears to be just a heavier copy of the electron.
In the quark sector, flavor-changing transitions occur with mixing defined by the CKM matrix.\cite{Kobayashi_1973}
The discovery of neutrino oscillations\cite{Fukuda_1998, Ahmad_2001, Ahmad_2002} is direct evidence that lepton flavor too is not a conserved quantity.
And yet similar flavor mixing has never been observed in the last class of elementary fermions, the charged leptons: the electron ($e$), the muon ($\mu$), and the tau ($\tau$).
Any observation of charged lepton flavor violation (CLFV) would be direct evidence of new physics\footnote{Neutrino oscillations present a loophole through which CLFV can occur, see for example the discussion in Ref.~\refcite{calibbi2017charged}. But as a neutral current flavor-changing process it is suppressed by a GIM-like mechanism\cite{Glashow_1970} making its predicted rate vanishingly small, $<10^{-50}$. It is no surprise then that this process has never been observed, and the statement stands: any experimental observation must be arising from new physics enhancements.}.

Searches for CLFV are very well motivated and share connections with areas of intense experimental and theoretical interest. 
The non-universal lepton interactions suggested by the $B$ meson anomalies are closely associated with violation of lepton flavor conservation.\cite{Glashow_2015}
And the same effective operators that mediate CLFV could have a flavor-conserving component that gives rise to the muon $g$-2 discrepancy, as well as to leptonic electric dipole moments. 
Finally, as the charged counterpart to neutrino oscillations, CLFV measurements can inform BSM models of the neutrino mass generation mechanism\cite{de_Gouvea_2013} and are synergistic with neutrino-less double beta decay experiments.

A rich global program is exploring this promising field, with complementary searches in different channels that hold the potential to elucidate the CLFV mechanism.
Searches for rare CLFV processes using muons are especially powerful and offer the best combination of new physics reach and experimental sensitivity, due to the availability of intense muon sources and their relatively long lifetime.
Of the many expansive reviews on the subject\cite{Marciano_2008,Mori_2014,Kuno_2001,Gorringe_2015}
we highlight especially the excellent introductory overviews of Bernstein and Cooper\cite{Bernstein_2013} and  Calibbi and Signorelli.\cite{calibbi2017charged}

There is a long history to experimental searches for CLFV between the first two families, where the highest sensitivity has been achieved. 
These searches began relatively soon after the discovery of the muon and have been closely connected to the evolution of our understanding of the SM flavor structure.
Limits on muon-to-electron transitions provided confirmation that the muon and electron neutrinos are separate particles\cite{Pontecorvo_1959} and gave rise to the concept of families of leptons, while suggesting a law of conservation of their flavor.\cite{Feinberg_1961}
Improvements in experimental sensitivity have largely been driven by the availability of more intense muon sources, from cosmic ray muons in the late 1940's to modern accelerator-based beams.
The current limits on the three most important CLFV muon modes are given below, all quoted at the $90\%$ confidence level (CL):
\begin{itemlist}
 \item In the $\mu^{+} \rightarrow e^{+} \gamma$ decay the branching ratio has been constrained to BR$(\mu^{+} \rightarrow e^{+} \gamma) < \num{4.2e-13}$.
Set by the MEG experiment\cite{MEG_2016} which collected data between 2008-2013 at PSI, this is the most stringent limit on any CLFV decay to date. 
Using a continuous beam to stop $7.5 \times 10^{14}$ muons on target, the experiment searched for coincidences between a positron and a photon emitted back to back and each with energy equal to half a muon mass.

\item The present limit BR$(\mu^{+} \rightarrow e^{+} e^{-} e^{+}) < \num{1.0e-12}$ is set by the SINDRUM collaboration.\cite{Bellgardt_1987}
The signature of this process consists of two positrons and one electron originating from a common vertex and with a total energy consistent with the mass of the muon.

\item The coherent muon to electron conversion in the field of a nucleus, $\mu^{-} N\rightarrow e^{-} N$, offers a cleaner experimental signature: here the only decay product to be detected is a monochromatic electron. 
Currently the best limit comes from the SINDRUM-II experiment at PSI, which stopped \num{4.4e13} muons on a  gold target. The limit on the conversion rate (normalized to muon captures, see Eq.~\ref{eqn:R_mu_e}) in muonic gold is\cite{SINDRUMII_2006} $R_{\mu e} \text{(Au)} < \num{7e-13}$.

\end{itemlist}

In the first two channels where a coincidence between particles must be identified experimentally, backgrounds arise when events accidentally have the correct topology and energies but the detected particles originate from uncorrelated mother processes.
This accidental background scales with the second or third power of the beam intensity and is likely the ultimate limitation for these channels.
The conversion process on the other hand offers significant advantages that allow powerful next-generation searches in the Mu2e\cite{Mu2e_2015} and COMET\cite{COMET_2009} experiments\footnote{
The COMET experiment at J-PARC employs similar design concepts to the ones presented here for Mu2e. COMET will adopt a staged approach and is expected to eventually reach similar sensitivity as Mu2e. Both experiments are under construction as of this writing.}
to take advantage of increased beam rate and reach sensitivities better than $10^{-16}$.
Details of this channel are discussed in the following section.

\subsection{Muon to electron conversion: The Mu2e search}

The Mu2e experiment at Fermilab\cite{Bernstein_2019} will search for the conversion of a muon to an electron in the field of a nucleus, $\mu^{-} N \rightarrow e^{-} N$.
There are no neutrinos in the final state so this is no standard muon decay; it is an interaction that has never been observed, which violates charged lepton flavor and signifies new physics.

The conversion is coherent, with the muon recoiling off the entire nucleus under two-body decay kinematics. 
The outgoing electron is monoenergetic with energy slightly less than the muon rest mass, $m_{\mu} \approx$ \SI{105.66}{\MeV}, after accounting for the muonic atom binding energy and the nuclear recoil. 
Mu2e has selected aluminum as the stopping material; for reasons that will be examined later it constitutes an excellent choice given the requirements to separate the signal from backgrounds.
The characteristic energy of the conversion electron (CE) on aluminum is\cite{Czarnecki_2011}
\begin{equation}
E_{\mu e} \text{(Al)}= m_\mu - E_\text{b} - E_\text{rec} = \SI{104.97}{\MeV} ,
\label{eqn:Ee}
\end{equation}
where $E_\text{b} $ is the binding energy of the muonic atom and $E_\text{rec}$ is the nuclear recoil energy.

This monoenergetic electron is the distinctive signature of CLFV that Mu2e will search for.
The nuclear recoil satisfies energy and momentum conservation without any other decay products in the final state.
Therefore this search is practically free from accidental coincidences\footnote{
The reconstruction errors described in Sec.~\ref{sec:recon_errors} may be construed as arising from accidental coincidences, but are a far smaller concern.}
, a limiting background contribution in other channels.
Furthermore the CE energy $E_{\mu e}$ is well above the maximum electron energy from stationary muon decay $E_e \approx m_\mu /2 = \SI{52.8}{\MeV}$,
separating the signal cleanly from the vast majority of electrons from muon decays\footnote{
The tail of this background spectrum is still an important consideration. It will be discussed in Sec.~\ref{sec:DIO}.}.

The conversion rate in the field of a nucleus is quoted relative to the rate of ordinary muon capture on the nucleus: 
\begin{equation}
R_{\mu e} = \frac{\mu^{-} \text{A(Z,N)} \rightarrow e^{-} \text{A(Z,N)}} {\mu^{-} \text{A(Z,N)} \rightarrow \nu_{\mu} \text{A(Z-1,N)}}
\label{eqn:R_mu_e} ~~,
\end{equation}
where the ratio formulation cancels uncertainties regarding the muon wavefunction and its overlap with the nucleus.
The current limit on the conversion rate is \num{7e-13} (90\% CL), set by the SINDRUM-II experiment which stopped \num{4.4e13} muons on a gold target.

The Mu2e experiment proposes to measure the muon to electron conversion rate on aluminum with a dramatic 4 orders of magnitude improvement, reaching sensitivity of better than $10^{-16}$ (90\% CL).
At that sensitivity Mu2e will probe new physics at effective mass scales up to \SI{10000}{\tera\eV}.\cite{de_Gouvea_2013}
Care must be taken in translating that effective scale to the direct mass reach of a high energy machine, but it is clear that in many scenarios Mu2e has sensitivity far beyond what is possible at LHC or any future collider.
Thus Mu2e has real discovery potential over a wide range of well motivated new physics models.
If CLFV is observed for the first time, complementarity between the muon channels can elaborate on the nature of the underlying mechanism.

In the following section we discuss the concepts that drive the design of Mu2e to achieve the proposed sensitivity improvement.

\section{Driving Concepts for a Next-Generation CLFV Search}

Mu2e aims to achieve an impressive four orders of magnitude improvement on $R_{\mu e}$ with respect to the current limit from SINDRUM-II.
The fundamental design goals required to achieve this sensitivity can be distilled in the following:
\begin{itemlist}
    \item The muon intensity must be increased by four orders of magnitude relative to previous experiments.
    Improvements in sensitivity have always been spurred by increased muon intensity.
    A solenoid muon capture system with a graded field, based on the design proposed by Dzhilkibaev and Lobashev\cite{Djilkibaev_MELC_1996}, will generate the most intense muon beam ever created. 
    
    \item A detector system is needed 	that can identify with high resolution the characteristic signature of CLFV, the conversion electron with momentum near \SI{105}{\MeVperc}.
    The detector must handle the high rates associated with the increased beam intensity.
    A clever design is required to make the detector blind to most background events and reduce occupancy.

    \item All potential sources of backgrounds must be suppressed.
    Since a single CE event is a powerful signal of new physics, Mu2e is aiming for extreme control of all background sources to reduce the expected number of background events to less than one over the entire lifetime of the experiment.

\end{itemlist}

\begin{figure}[t]
\centerline{\includegraphics[width=\columnwidth]{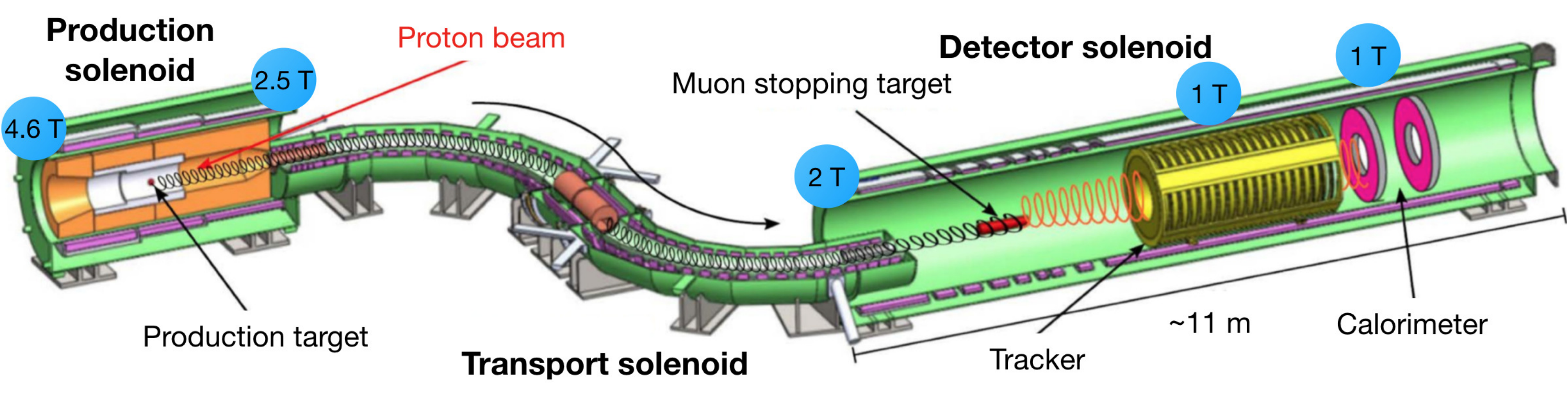}}
\vspace*{10pt}
\caption{The Mu2e experimental apparatus. 
Also drawn are the helical trajectories of particles in the solenoid fields. The varying field amplitude is listed in several locations.
\protect\label{fig:apparatus}}
\end{figure}

The Mu2e experimental apparatus that was developed along these conceptual drivers is shown in Fig.~\ref{fig:apparatus}.
In the following three sections we delve deeper into each one of these design goals and how the Mu2e design aims to achieve them.

\section{The Mu2e Solenoids: Generating the World's Most Intense Muon Beam}

The Mu2e solenoid system is the most innovative, essential, and technically challenging component of the experiment. 
It consists of approximately {\SI{75}{km}} of NbTi superconducting cable stabilized with high conductivity aluminum. 
The generated fields are required to efficiently capture charged pions from the production target and transport negatively charged secondary muons to the stopping target.
Muons of sufficiently low momentum must be selected such that a significant fraction can be stopped in a thin target, while transmission of other particles is minimized.

To aid in the efficient capture and muon beam creation the solenoid system creates a continuously graded magnetic field, from 4.6 down to \SI{1}{Tesla}.
The gradient also suppresses backgrounds by preventing the local trapping of particles as they traverse the muon beamline, and by pitching beam backgrounds forward and out of the detector acceptance.
The field then remains nearly uniform at the detector region to facilitate momentum analysis of conversion electrons.

The solenoid system consists of three functional units, referred to as the Production Solenoid (PS), the Transport Solenoid (TS), and the Detector Solenoid (DS).
The fringe field from each impacts the field in adjacent units and significant forces develop, so the solenoids have to be designed and operate as a single integrated magnetic system.
The three solenoids are presented in this section after an overview of the proton beam delivered by the Fermilab accelerator complex.

\subsection{FNAL proton beam}

To produce its intense beam of low energy muons, the experiment requires a high intensity proton beam with a pulsed time structure.
The Fermilab accelerator complex can provide this beam with a macroscopic duty factor of about 30\%, simultaneously with operations and beam delivery to the NOvA experiment.
Protons with a kinetic energy of \SI{8}{GeV} are transferred from the Fermilab Recycler Ring to the Delivery Ring in \SI{2.5}{\MHz} bunches;
proton pulses are then resonantly extracted to Mu2e with a pulse spacing of \SI{1695}{\ns}, equal to the DR revolution period. 
Each pulse delivers about \num{3.9e7} protons at \SI{8}{\GeV} to the Mu2e production target. 
A total of \num{3.6e20} protons on target are required to reach the Mu2e target sensitivity. 
That should correspond to $\sim$\num{7e17} stopped muons, four orders of magnitude more than was accomplished in SINDRUM-II.

\subsection{Production Solenoid}

The Production Solenoid (PS) houses the Mu2e production target in a graded field varying smoothly from 2.5 to \SI{4.6}{T} to maximize secondary beam collection. 
The \SI{8}{\GeV} pulsed proton beam from the Fermilab accelerator enters the PS from the low-field side, moving in the direction of increasing field strength towards the production target, where it interacts.
The graded field collects the backward-produced pions\footnote{
Some pions produced in the forward direction at high angles relative to the solenoid axis will also be reflected backwards by the field gradient.}
, steadily increasing their pitch and accelerating them in helical trajectories towards the lower field of the Transport Solenoid. 
Most of the soft pions decay into muons within a few meters.

The production target is made from tungsten, a high-Z material to maximize pion production.
The target has a small physical profile, is suspended by spokes and is allowed to cool radiatively - a geometry that aims to minimize scattering and reabsorption of pions.
Vacuum here must be maintained to better than \SI{e-5}{Torr} to minimize tungsten oxidation and corrosion that reduce target lifetime.

The diameter of the warm bore of the PS is large enough to allow pions and muons within the acceptance of the Transport Solenoid to pass through unobstructed.
A heat and radiation shield, constructed from bronze, will line the inside of the PS to limit the heat load in the cold mass from secondaries produced in the production target and to limit radiation damage to the superconducting cable.
To repair displacement of atoms from the aluminum stabilizer lattice due to radiation damage, the system will be annealed once per year (during accelerator shutdowns) to allow thermal motion to repair the lattice.

\subsection{ Transport Solenoid }

\begin{wrapfigure}{r}{0.6\textwidth}
	\centering
	\vspace*{9pt}
	\includegraphics[width=0.56\columnwidth]{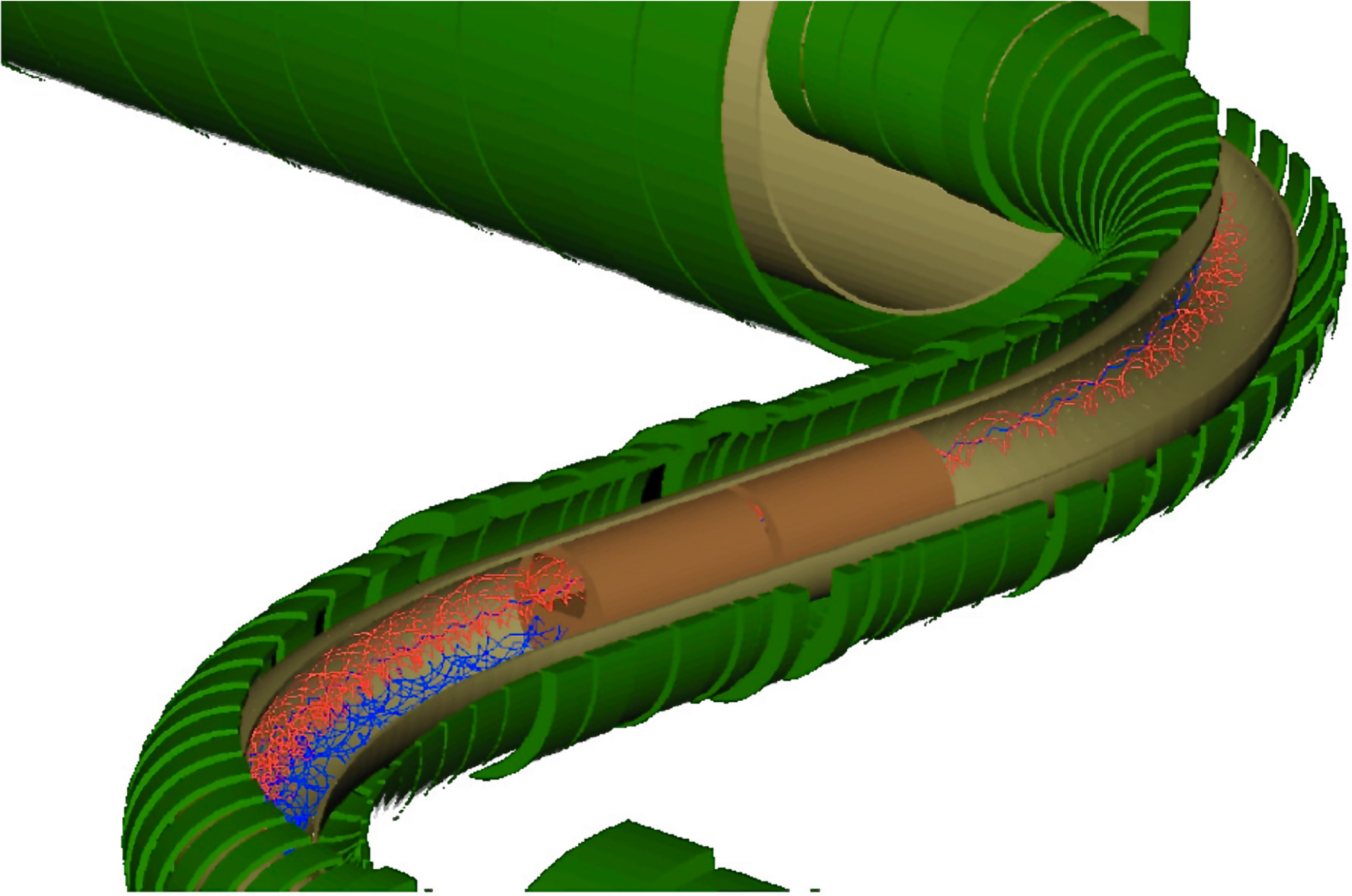}
	\caption{Positive (blue) and negative (red) particles receive an opposite vertical offset in the TS toroidal bend. In normal operation a central collimator only allows negative muons through to the stopping target.
	}
	\protect\label{fig:TS_selection}
\end{wrapfigure}

The Transport Solenoid (TS) has a characteristic S-shape that eliminates line-of-sight neutral particles from the Production Solenoid.
Its field efficiently collects and transmits low energy negatively charged particles from the PS to the stopping target, while directing high energy and positively charged particles onto absorbers and collimators. 
Depicted in Fig.~\ref{fig:TS_selection}, the selection relies on the toroidal field on the first bend which induces a sign- and momentum-dependent vertical dispersion; this couples with a vertically displaced aperture on the central collimator to absorb most positive particles and allow low momentum negative particles to pass through.
The second toroidal bend then largely undoes the vertical offset and returns the now mostly negative beam close to the solenoid axis.
To accommodate detector calibration procedures the central collimator can be mechanically rotated to instead allow positive particles through.

\subsection{ Detector Solenoid }

The Detector Solenoid (DS) is a large, low field magnet that houses the muon stopping target and the detectors that will identify and analyze conversion electrons. 
The efficient capture and transport to this point results in approximately \num{0.005} generated muons per proton-on-target.
About 40\% of these muons will be stopped at the aluminum target and these tend to have lower momentum, with a peak around \SI{35}{\MeVperc}.
This corresponds to a rate of 10$^{10}$ stopped muons per second.

The aluminum stopping target must be massive enough to stop a significant fraction of the incident muon beam, but it must also allow conversion electrons from stopped muons to emerge without having their momentum corrupted by energy loss and straggling, as that would make separation between signal and background more difficult.
The stopping target design is a balance between these two competing requirements. 
It consists of a series of 34 thin foils, each \SI{100}{\um} thick and composed of $>99.99\%$ pure aluminum, arranged coaxially along the Detector Solenoid axis.
The field is axially graded near the stopping target, linearly decreasing from \SI{2}{T} at the entrance of the DS down to \SI{1}{T} over roughly five meters. 
This gradient nearly doubles the detector acceptance for conversion electrons while also removing backgrounds from beam electrons by pitching them forward.

\bigskip
At this point we interrupt the description of the apparatus to describe processes that occur in the stopping target, and especially the muon decay in-orbit {\textendash} a background source that is a significant driver of the spectrometer design.

\subsubsection{Processes at the stopping target}

When muons are stopped on an aluminum atom they displace an electron and then very rapidly (within $\sim$\SI{1}{fs}) cascade down to the lowest 1S atomic orbital, emitting X-rays at characteristic energies. 
There are three possibilities on the fate of a muon stopped in orbit:

\begin{enumerate}
\item Some of these stopped muons may convert to an electron, in the process that Mu2e searches for which could yield the first observed signal of CLFV.
We know however that this process must be very rare, with $R_{\mu e}$ < \num{7e-13}.
The other two possibilities are thus far more likely to occur.

\item Roughly 60\% of stopped muons on aluminum will be captured in the nucleus. 
This is not surprising given the smallness of the Fermi radius of the muonic atom, $\sim$ \SI{20}{fm}. 
The muonic wavefunction has significant overlap with the nucleus and capture is common. 
This is the process that appears in the denominator of Eq.~\ref{eqn:R_mu_e} and is used for rate normalization.

\item The remaining 40\% of stopped muons will undergo standard decay {$\mu \rightarrow e \nu_\mu \bar{\nu}_e$} while in orbit around the nucleus.
\end{enumerate}

The lifetime of a free muon is $\tau_\text{free} = \SI{2.2}{\us}$, but a muonic atom is limited by both capture and decay probabilities: 
$\tau = 1/(\Gamma_\text{capture} + \Gamma_\text{decay})$.
The capture probability is nucleus-dependent as the overlap of the muon wavefunction varies for different elements.
The muonic aluminum lifetime is measured to be $\tau = \SI{864}{\ns}$.\cite{Measday_2001,Suzuki_1987}
For reasons that will be discussed when we present the experiment's time structure in Sec.~\ref{sec:time_structure}, the lifetime of muonic aluminum is an excellent fit to Mu2e requirements and a central reason for the selection of aluminum as stopping material.

The muon decay in-orbit (DIO) process is a significant background consideration which drives the spectrometer requirements and design. 
We discuss it in the following paragraph before going further.

\subsubsection{Muon decay in orbit}
\label{sec:DIO}

The standard weak decay of the muon with associated neutrinos, $\mu \rightarrow e \nu_\mu \bar{\nu}_e$, is well understood.
For free muons this decay was studied first by Michel\cite{Michel_1950} and the spectrum of the decay electron is often referred to as the `Michel spectrum'.
4-momentum conservation demands that the neutrinos take away at least half of the available energy, therefore the maximum energy of the outgoing electron (neglecting neutrino masses) is $E_\text{max} = (m_{\mu}^2 + m_{e}^2 ) / (2 m_{e}^2 ) = \SI{52.8}{\MeV}$.

\begin{figure}[t]
\centerline{\includegraphics[width=\columnwidth]{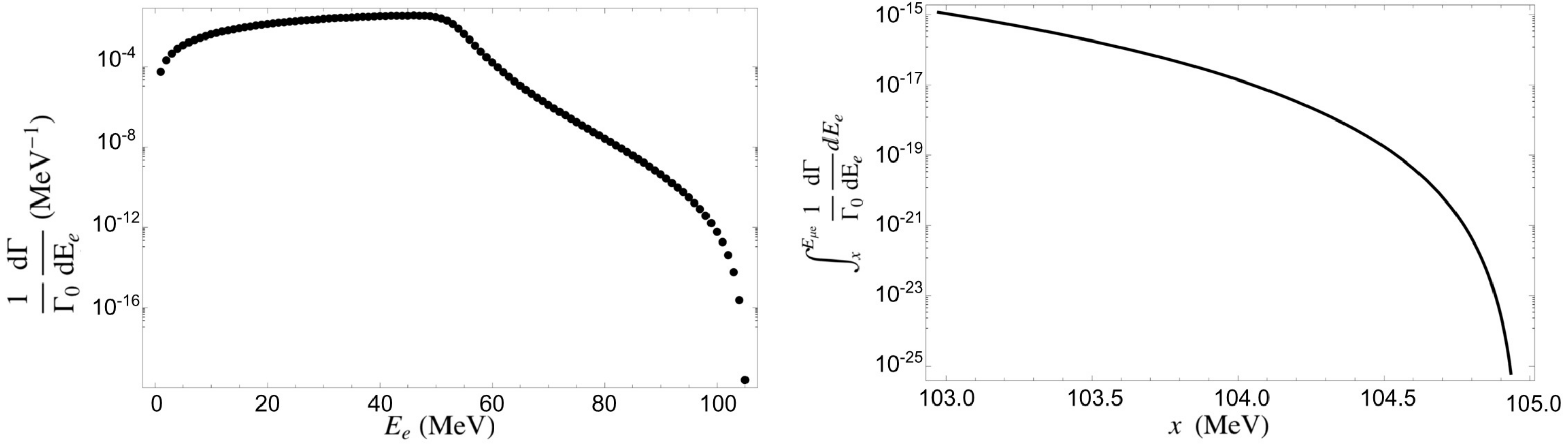}}
\vspace*{9pt}
\caption{Left: The DIO spectrum from Czarnecki \etal\cite{Czarnecki_2011}, logarithmic scale. 
Right: Total contribution to $R_{\mu e}$(Al) from DIO electrons with energy larger than $x$, as a function of $x$. 
To limit this background contribution the CE search must be initiated above a high enough energy threshold near $E_{\mu e}$.
\protect\label{fig:DIO}}
\end{figure}

$E_\text{max}$ is far smaller than the characteristic conversion electron energy $E_{\mu e} = \SI{104.97}{\MeV}$ from Eq.~\ref{eqn:Ee}, allowing a clean separation from background.
However the decay in-orbit (DIO) energy spectrum is modified in the field of the nucleus: the nuclear recoil allows 4-momentum conservation with the neutrinos taking vanishingly small energy, pushing the endpoint of the DIO spectrum near $E_{\mu e}$.
The modified spectrum is shown in Fig.~\ref{fig:DIO}, properly incorporating nuclear recoil and relativistic effects near the high energy endpoint.\cite{Czarnecki_2011}
Leading radiative corrections to the high energy tail have been calculated in Ref.~\refcite{Szafron_2016a}.
The long tail of the spectrum falls very rapidly near the CE energy due to the diminishing phase space available to produce neutrinos that carry almost no energy\footnote{
It is no surprise that the simulated signal-background spectrum (presented in Fig.~\ref{fig:signal_pseudoexperiment}) looks very similar to that of a neutrino-less double-beta decay search; the kinematics are very similar.}.
The DIO rate drops by more than 16 orders of magnitude near the endpoint relative to the Michel peak, allowing a search with the proposed Mu2e sensitivity.

Taking into account energy loss in material and detector resolution, the tail of the DIO component overlaps with the expected CE distribution and constitutes an irreducible SM background contribution to the Mu2e measurement.
This component arises from muons stopped at the target and therefore has the same timing signature as the signal.
The only way to separate the CE signal from an electron at the tail of the DIO distribution is a high resolution measurement of the electron momentum, to identify the difference in energy carried off by the two neutrinos.
The detector system that will perform this highly sensitive and critical measurement is described in the next section.

\section{The Mu2e Detector} \label{sec:detector}

The Mu2e detector system consists of a tracker followed and complemented by an electromagnetic calorimeter, located inside the Detector Solenoid downstream of the stopping target within a nearly uniform \SI{1}{Tesla} magnetic field.
It is designed to efficiently and accurately identify and analyze the helical trajectories of $\sim${\SI{105}{\MeV}} electrons in a high rate time-varying environment, while rejecting backgrounds from conventional processes.
A high precision momentum measurement is required to disentangle the CE signal from the DIO background.
The reconstructed width of the conversion electron energy peak, including energy loss and resolution effects, must be narrow enough to keep DIO backgrounds at an acceptably low level.
The detector is placed in a vacuum of \SI{e-4}{Torr} to minimize multiple scattering contributions to the momentum resolution, and to suppress potential backgrounds sources.

\subsection{Tracker}

The Mu2e tracker is designed to accurately determine the trajectory of $\sim${\SI{105}{\MeV}} electrons in a uniform 1 Tesla magnetic field in order to measure their momenta. 
Its core momentum resolution must be $<\SI{180}{\keVperc}$ for \SI{105}{\MeVperc} electrons, or 0.17\%, to provide separation from the high energy DIO spectrum.
High-side tails of the resolution especially need to be controlled since they could push a DIO electron of slightly lower energy to be misidentified as a conversion electron.
The tracker must present very low mass in the path of the electron, in order to minimize energy loss and smearing of the momentum measurement.
It must also be highly segmented to handle high rates that could lead to pattern recognition errors.

The detector technology that was chosen to meet these requirements is that of ``straw'' drift tubes.
Charged particles traversing a straw will ionize the gas inside the tube and generate a signal in the HV sense wire at its center.
After pre-amplification the signal is taken to a TDC (implemented in FPGA) for a measurement of the drift time that determines the distance of closest approach of the particle relative to the straw wire.
Each straw is instrumented for readout on both sides, allowing for a time difference measurement between the two sides which determines the position of the hit along the length of the straw with a resolution of about \SI{4}{cm}.
The straws are also instrumented with an ADC for a $dE/dx$ measurement that allows identification and removal of highly ionizing hits from protons. 
The information from all the straws intercepted by the particle is then combined and used in the reconstruction of its trajectory, which yields a measurement of its momentum with high precision.

\begin{figure}[b]
\centerline{\includegraphics[width=\columnwidth]{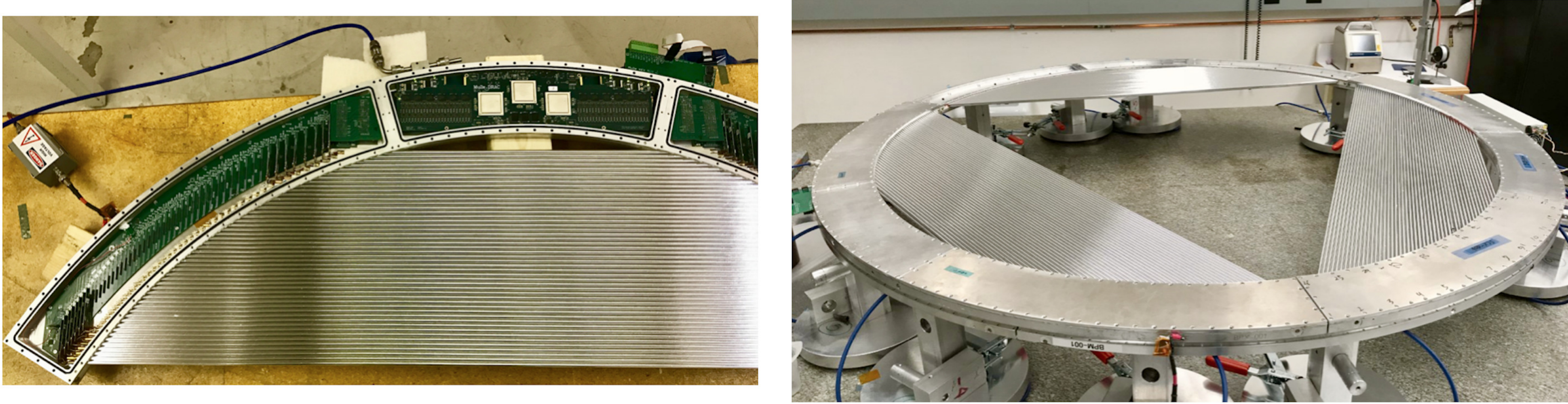}}
\vspace*{9pt}
\caption{Left: A prototype of the Mu2e tracker panel with exposed electronics boards (author's photo). Right: Three panels arranged in a $360\degree$ ring (photo courtesy of Yujing Sun). 
\protect\label{fig:panel}}
\end{figure}

The straw tubes are \SI{5}{mm} in diameter and aligned transverse to the DS axis. 
Minimization of material in the path of the particle is imperative to momentum resolution, so the straw wall thickness is only \SI{15}{\um}, consisting of two layers of spiral wound Mylar$^\text{\textregistered}$.
That includes metalization on the inside surface of the straw (\SI{500}{\angstrom} of aluminum overlaid with \SI{200}{\angstrom} of gold) so it can act as the cathode layer, and on the outside surface (\SI{500}{\angstrom} of aluminum) for additional electrostatic shielding and to reduce gas leak rate.
The anode sense wire inside each straw is \SI{25}{\um} of gold-plated tungsten.
With 80:20 Ar:CO$_2$ as the drift gas the sense wire will be operated near \SI{1400}{V}.
In total the tracker presents approximately only 1\% radiation length of material to the average CE.

Straws vary in active length from 334 to \SI{1174}{mm} and are supported at the ends in $120\degree$ holding fixtures referred to as panels, which reside outside the active detector region.
Each panel holds 96 straws mounted in two staggered layers to improve efficiency and help identify on which side of the sense wire a particle passed, which the signal from an individual straw can not resolve.
Prototypes of tracker panels are shown in Fig.~\ref{fig:panel}.
To minimize penetrations into the vacuum, digitization and readout will be performed on-board each panel via three FPGAs, with optical fiber readout. 
Three panels are required to cover a $360\degree$ ring, and each subsequent ring is rotated by $30\degree$ for better stereographic reconstruction.
The tracker will have $\sim$21,000 straws distributed across a length of \SI{3270}{mm}, in a design with high segmentation that will allow the detector to handle high rates.

\begin{figure}[b]
\centerline{\includegraphics[width=0.75\columnwidth]{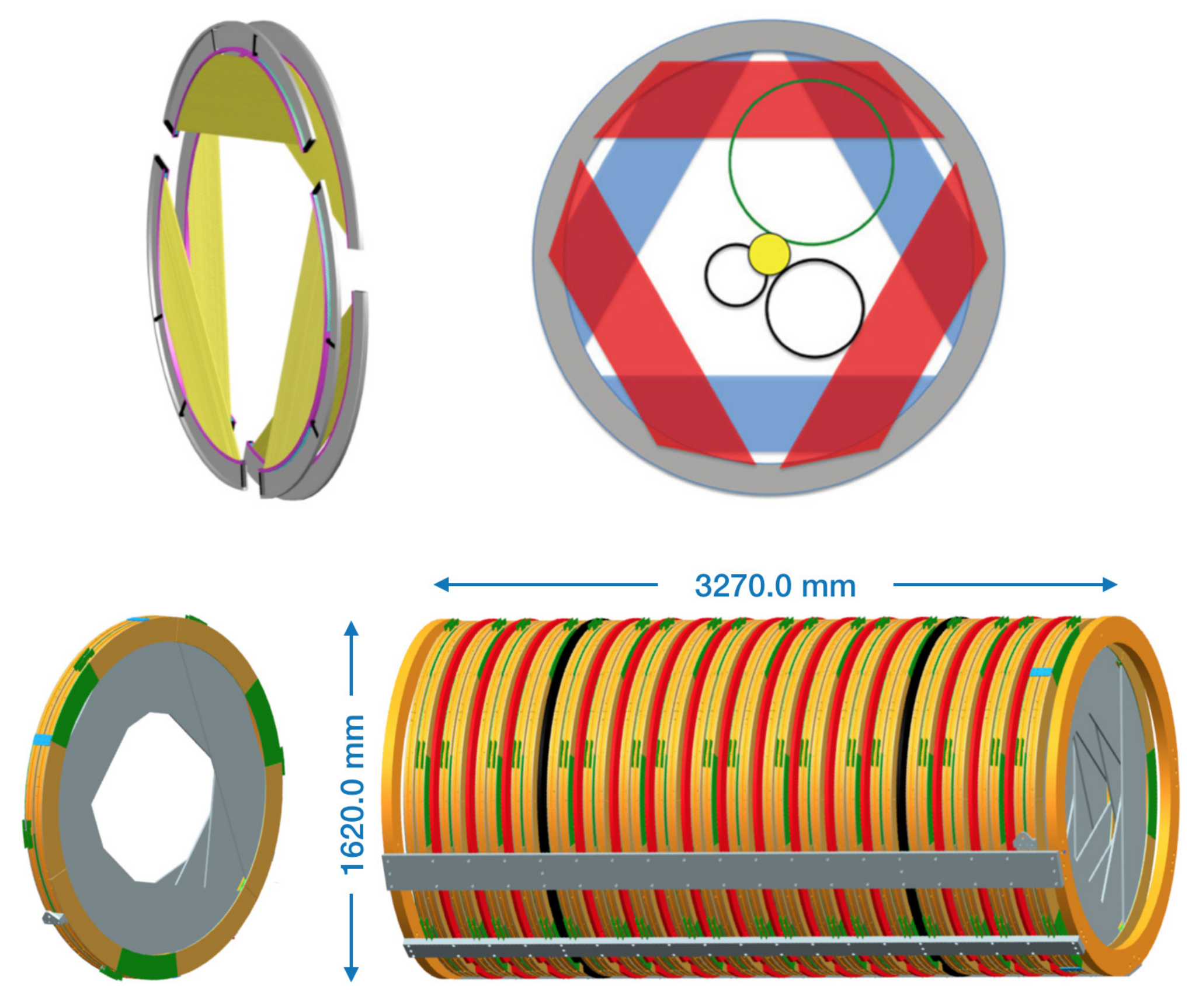}}
\vspace*{9pt}
\caption{Design of the Mu2e tracker. $120\degree$ panels are assembled into rings, successively rotated by $30\degree$. 
The annular design (top right, beam's eye view) makes the detector blind to the majority of DIO electrons and remnant beam (dark circles), while remaining sensitive to events with high transverse momentum (green circle).
A schematic of the assembled tracker is shown in the bottom right figure.
\protect\label{fig:tracker_design}}
\end{figure}

As shown in Fig.~\ref{fig:DIO} the muon decay in-orbit spectrum decreases steeply above $m_{\mu}/2$, a factor of two away from the endpoint.
This allows the opportunity for a detector design with good geometrical acceptance for signal and away from most of background events.
An annular design is employed with a hole near the solenoid axis where no detector material exists, making the detector blind to the vast majority of DIO electrons, the remnant beam, and other products from the stopping target.
All these particles would overwhelm the detector with high instantaneous occupancy and several adverse effects: increased dead time, more difficult pattern recognition and increased risk of backgrounds from reconstruction errors, and larger charge deposition on the straws. 
In the design shown in Fig.~\ref{fig:tracker_design} the tracker straws are located in larger radii, between $380 < r < \SI{700}{\mm}$ (where the radius $r$ is measured from the center of the muon beam), allowing the tracker to be fully efficient for electrons coming from the target with transverse momenta above \SI{90}{\MeVperc}.
This allows only $\mathcal{O}${(\num{e-12})} of the DIO events to produce reconstructable tracks.

The tracker will still be exposed to high rates, which generally vary with radius from center, distance from target, and time after the proton pulse.
On average the straw rate within the detector live window (defined in Sec.~\ref{sec:time_structure}) is $\sim${\SI{20}{\kHz/\cm\squared}}, with up to \SI{5}{\mega\Hz} rate on individual straws.
The annular design reduces the occupancy, rate, and radiation damage on the straws to a manageable level.

\begin{figure}[h]
\centerline{\includegraphics[width=\columnwidth]{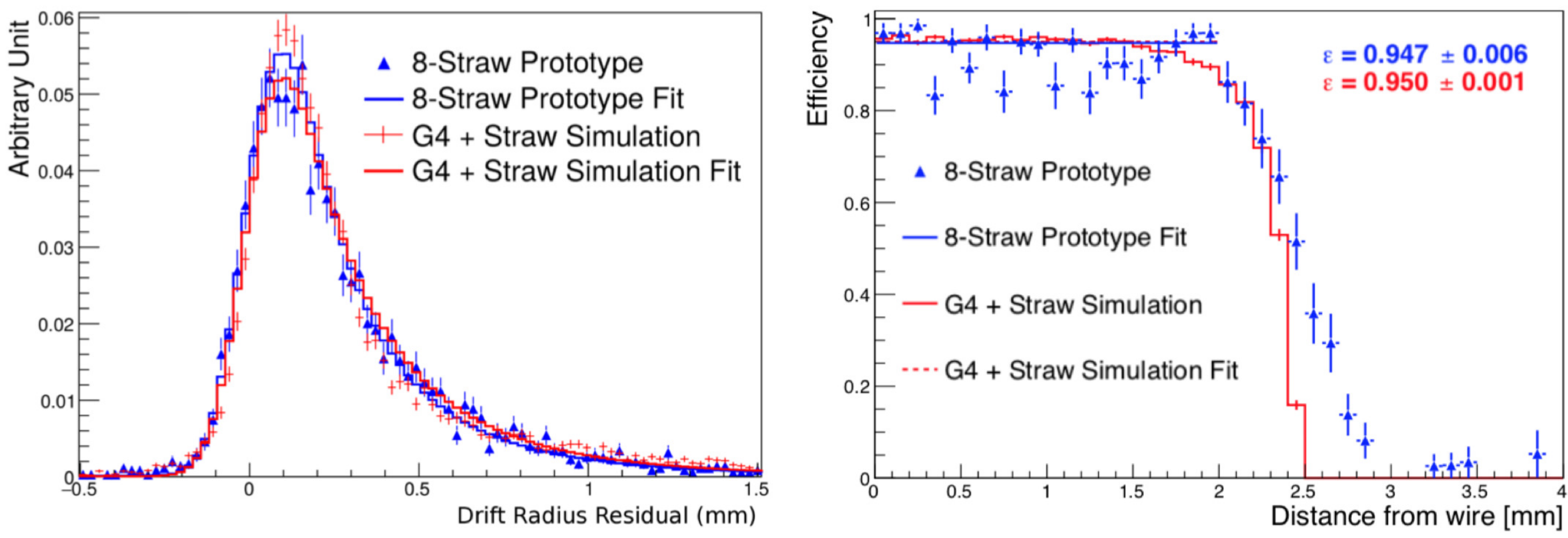}}
\vspace*{9pt}
\caption{Transverse resolution (left) and efficiency (right) for cosmic ray events on an 8-straw tracker panel prototype. Data shown in blue, Monte Carlo simulation in red. From Bonventre.\cite{Bonventre_2019}
\protect\label{fig:tracker_prototype_plots}}
\end{figure}

The performance of the tracker design has been characterized with an 8-channel panel prototype,\cite{Bonventre_2019} with results from cosmic ray events shown in Fig.~\ref{fig:tracker_prototype_plots}. 
The prototype demonstrates transverse resolution with a FWHM of \SI{283}{\micro\m}, and longitudinal resolution with a core width of \SI{43}{\mm}, meeting the experimental requirements for the tracker. 
An efficiency of 95\% is achieved for events closer than \SI{2}{\mm} from the sense wire.
The data is nicely in agreement with a full Geant4-based simulation.

\subsection{Calorimeter}

The tracker is followed downstream by a calorimeter system, which consists of pure cesium iodide (CsI) crystals read by silicon photo-multipliers (SiPMs). 
The calorimeter provides redundant energy, position, and timing information that complements the tracker, providing particle identification and background rejection capabilities, as well as a standalone trigger for high energy electrons. 
High hit rates in the tracker may cause pattern recognition errors that add tails to the resolution function, potentially misidentifying a DIO track as a CE-consistent event. 
%, and in some cases even in a pattern consistent with a CE where there is none.
By extrapolating the fitted trajectory and comparing with clusters at the calorimeter, events can be confirmed and the momentum resolution improved.
Calorimeter clusters may also seed track finding with increased efficiency, reducing the combinatorial background by using only straw hits within $\sim\SI{50}{ns}$\footnote{The time range corresponds to the maximum drift time in a straw.} from the detected cluster.

\begin{figure}[t]
\centerline{\includegraphics[width=0.9\columnwidth]{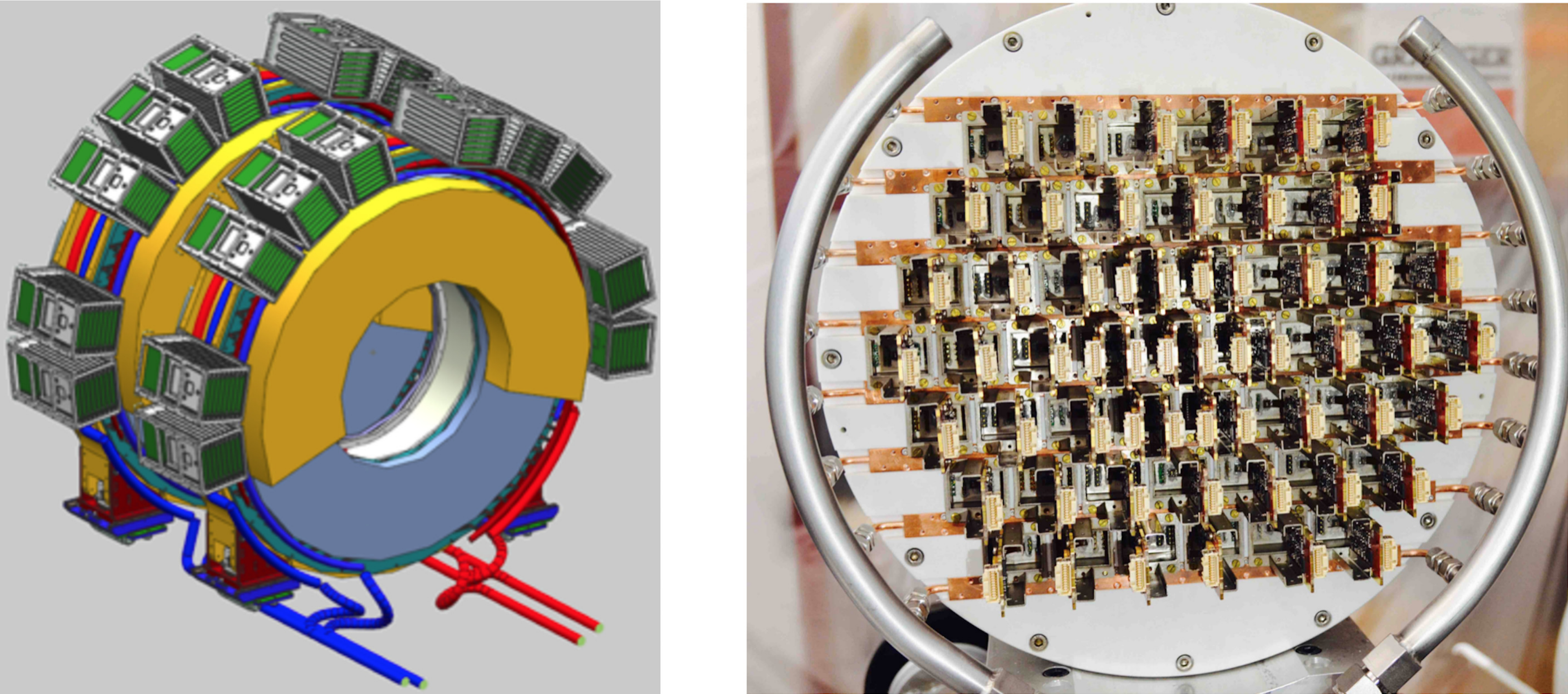}}
\vspace*{9pt}
\caption{Left: CAD layout of the two disks of the calorimeter. Right: a large size prototype built with pre-production components. From Atanov \etal\cite{Atanov_2018}
\protect\label{fig:calo_combined2}}
\end{figure}

The Mu2e electromagnetic calorimeter\cite{Atanov_2016} consists of two annular disks composed of a total of \num{1348} undoped CsI scintillating crystals.
The dimension of a crystal is $3.4 \times 3.4 \times$ {\SI{20}{\centi\metre\cubed}} for a total depth of 10 $X_0$, enough to contain the CE shower. 
Each crystal is read out by two custom large area UV-extended SiPMs. Like the tracker, the calorimeter also employs an annular design with a hole in the center to minimize interactions with the remnant beam and the beam flash.
%%% the second part of this sentence (after the colon) should be worded better:
The distance between the calorimeter disks is \SI{70}{\cm}, optimized to maximize acceptance: the helical trajectory of a CE reconstructable at the tracker has a good probability of intercepting the second disk if it passes through the hole of the first.
The calorimeter layout is shown in Fig.~\ref{fig:calo_combined2}.

The calorimeter energy, position, and timing information must provide confirmation of the tracker measurement, to help reject backgrounds from spurious combinations of straw hits. 
Hence the position and timing resolution must be comparable to the error of the extrapolated track, including multiple scattering at the tracker.
The energy and timing resolution must also allow particle identification for separation of CE candidate events from cosmic ray muons.  
This set of criteria defines the calorimeter resolution requirements: better than $\sigma_E /E < 10\%$, $\sigma_{x,y} < \SI{1}{\cm}$, and $\sigma_t < \SI{500}{ps}$ for a \SI{105}{MeV} electron must be achieved.

\begin{figure}[t]
\centerline{\includegraphics[width=\columnwidth]{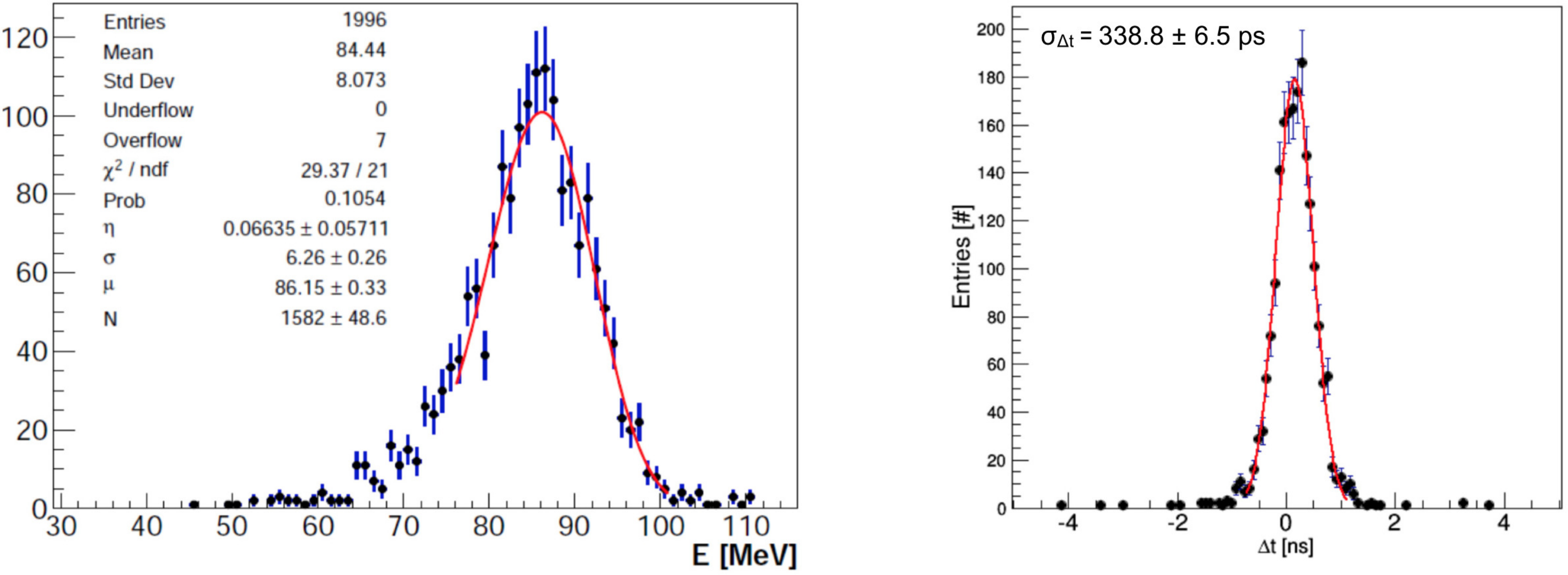}}
\vspace*{9pt}
\caption{Results for energy resolution (left) and preliminary time difference between the SiPMs reading out a single crystal (right) for \SI{100}{MeV} electrons on a small scale prototype of the Mu2e calorimeter, at an impact angle of $50\degree$. 
From Atanov \etal \cite{Atanov_2018}.
\protect\label{fig:calo_prototype_plots}}
\end{figure}

The calorimeter performance has been characterized with a small scale prototype of 51 crystals at the Beam Test Facility in Frascati.\cite{Atanov_2018}
Results for \SI{100}{MeV} electrons at an impact angle of $50\degree$, near the average experimental condition for CE, are shown in Fig.~\ref{fig:calo_prototype_plots}.
The energy resolution is found to be $\sigma_E /E \sim 7\%$, and a preliminary estimation of $\sigma_t \sim \SI{170}{ps}$ is found with  non-final electronics on the average of the two SiPMs reading the crystal with the highest energy deposit. 
The results are shown to be well within requirements, and also in good agreement with simulation.

\section{Backgrounds and Suppression Strategies}

Searching for a process that has never before been observed, Mu2e is aiming for an expectation of less than 1 background event over the lifetime of the experiment.
While the experimental signature is clean, at the proposed sensitivity several small background contributions become important.
Suppressing and identifying all possible sources of background is an important driver of many aspects of the Mu2e design.
The three most important background sources in Mu2e are:
\begin{itemize}
    \item \textit{Intrinsic processes} that are associated with the stopped muons and scale with beam intensity, including muon decay in orbit (DIO) and radiative muon capture (RMC).
    \item \textit{Prompt processes} that occur shortly after the arrival of the proton pulse, including radiative pion capture (RPC), muon and pion decay in flight (DIF), and beam electrons.
    \item \textit{Cosmic ray-induced processes}.
\end{itemize}

Other backgrounds arise from delayed processes due to particles that spiral slowly down the beamline, such as antiprotons;
    and reconstruction errors at the detector induced by high occupancy from conventional processes.
In this section we discuss the different background classes and mitigation strategies adopted by Mu2e.

\subsection{Intrinsic backgrounds} 

Intrinsic (or muon-induced) backgrounds are associated with stopped muons and therefore have the same time signature as our search.
Our only recourse to provide separation from a conversion signal is a high precision momentum determination.
The most important intrinsic source is muons stopped at the aluminum target that decay in orbit (DIO) around the nucleus through ordinary weak decay.
This is the process discussed in Section~\ref{sec:DIO}, and in Section~\ref{sec:detector} we presented the detector design which optimizes resolution to separate the conversion signal from this background.

Another important process is radiative muon capture (RMC), $\mu^- + \text{Al} \rightarrow e^- + \bar{\nu_e} + \nu_\mu + \text{Mg} + \gamma$.
This is an intrinsic source of high energy photons that can convert to an electron-positron pair in the stopping target or other surrounding material, producing an electron near the conversion electron energy.
Consideration for this process was another reason for selecting aluminum as the stopping material:
the nuclear mass difference between Al and Mg is \SI{3.11}{\MeV}, placing the RMC endpoint well below the CE energy.
Thus the RMC background is well separated from the signal, given the momentum resolution of the detector, and it is unlikely to be a significant source of background\footnote{
It is still a consideration however that RMC events will distort the DIO spectrum since they overlap in the 80$-$\SI{100}{\MeVperc} range. 
The charge-symmetric detector configuration will allow detection of both RMC $e^-$ and $e^+$, making it possible to disentangle the background processes in situ.}.

\subsection{Prompt backgrounds}

A prompt event is defined as one that occurs shortly (within a few hundred ns) after the proton beam interaction at the production target.
The most relevant example in this class of backgrounds is radiative pion capture (RPC) $\pi^{-} N \rightarrow \gamma N'$, where $N'$ is an excited nuclear state. 
This process occurs promptly as the pion stops in the aluminum target. 
The radiated photon (whether on-shell or virtual) can then convert into an electron-positron pair. 
The pion mass allows a high enough endpoint for the photon energy, so that an asymmetric conversion can result in an electron near \SI{105}{\MeV} which mimics the CE signal. 
Other examples of prompt background events include muons and pions decaying in-flight with enough energy, and beam electrons near the conversion energy that scatter in the target.

The SINDRUM-II experiment used beam counters to tag and veto prompt backgrounds, for which the relevant timescale is the pion lifetime of \SI{26}{ns}. 
The PSI beam was a continuous stream of \SI{20}{ns} bursts, not allowing separation of this background from the measurement. 
This was the ultimate limitation of the SINDRUM-II method which could not go to higher beam intensity without being overwhelmed by this background.
Mu2e aims to increase its sensitivity, and for that its beam intensity, by four orders of magnitude. 
Suppression of prompt background sources poses very specific requirements on the proton beam time structure, which is described in the following section.

\subsubsection{The Mu2e time structure}
\label{sec:time_structure}

\begin{figure}[t]
\centerline{\includegraphics[width=\columnwidth]{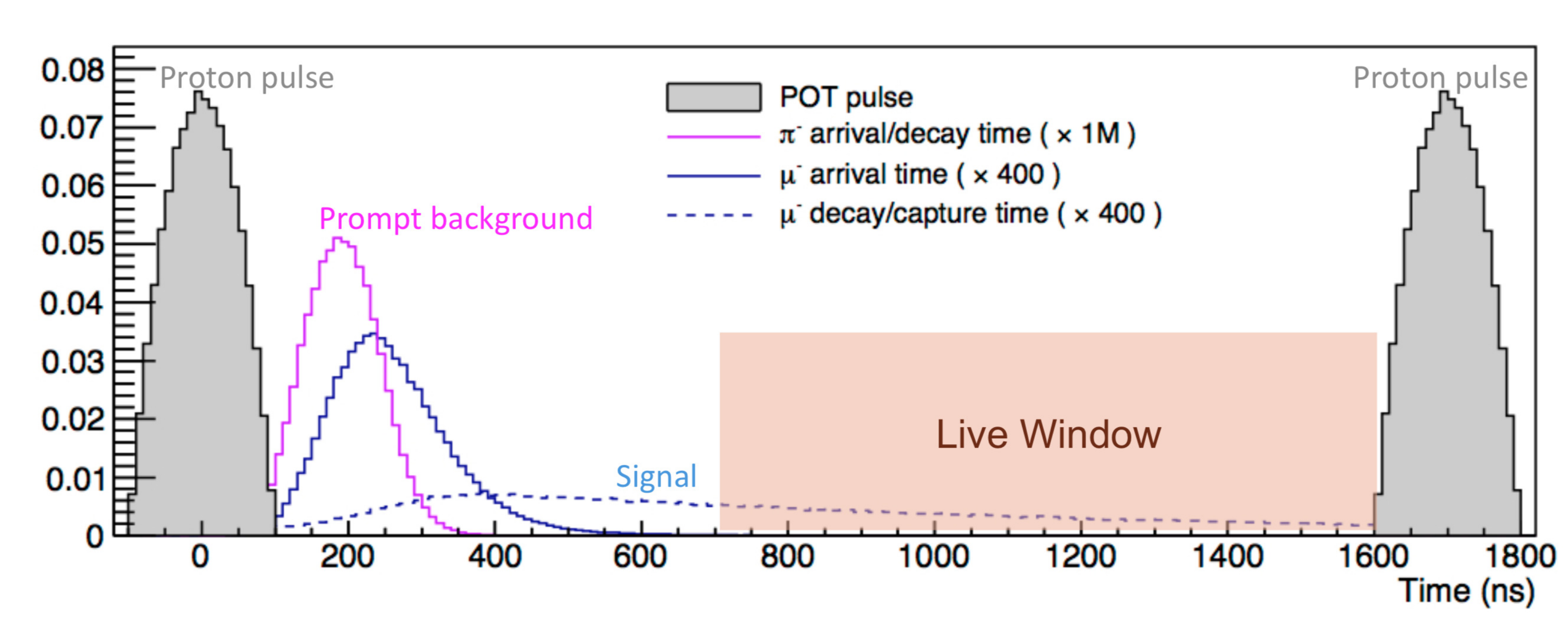}}
\vspace*{9pt}
\caption{The Mu2e time structure. 
A pulse of \num{3.9e7} protons arrives at the production target every \SI{1695}{\ns}. 
The live window for the conversion search is only initiated about \SI{700}{\ns} later, after the probability of observing a background event from a prompt process has become negligible.
\protect\label{fig:time_structure} }
\end{figure}

In Mu2e the main concept to suppress backgrounds from prompt processes is to take advantage of their short lifetime and simply wait for them to vanish before initializing a search for the conversion signal.
The time structure that accomplishes that is shown in Fig.~\ref{fig:time_structure}.
Pion-related and other prompt backgrounds decay soon after the arrival of the proton pulse, characterized by the the short \SI{26}{\ns} pion lifetime and short time to transit the beamline. 
The live time for measurement then begins after about \SI{700}{\ns}, only after enough pions have decayed and prompt events are negligible.
The relatively long lifetime of muonic aluminum at \SI{864}{\ns} is the main reason for the selection of aluminum as stopping material: it allows a significant population of stopped muons after \SI{700}{\ns} that can convert within the live window.
Thus the CE signal is well separated in time from the beam flash and prompt processes.

The live window extends until the arrival of the next pulse. 
Notice that the \SI{1695}{\ns} spacing between beam pulses, though fixed by the revolution period of protons at the Fermilab Delivery Ring, is a very good fit to Mu2e requirements. 
At roughly twice the lifetime of muonic aluminum, it allows enough time for the CE search after the delay to suppress prompt backgrounds, and sends a new pulse when the signal probability has diminished significantly.
This scheme is also fully compatible with operation of the Fermilab neutrino program.

Here we remind the reader of the graded field of the solenoids and the requirement to avoid trapping particles, the reason for which should now become clear.
Magnetic trapping of particles between local maxima would delay their arrival at the stopping target, allowing a prompt background event within the observation window.
The same could occur if a proton arrives at the production target outside the main proton pulse.
The resonant extraction process from the Delivery Ring suppresses protons between pulses, and a fast AC dipole is employed to further sweep clean the inter-pulse beam.\cite{Mott_2016}
The ratio of out-of-time beam to protons in the main pulse (referred to as beam extinction) is expected to be better than 10$^{-10}$ while transmitting $\sim$99.7\% of the in-time beam.

\subsection{Cosmic rays}

\begin{figure}[t]
\centerline{\includegraphics[width=0.97\columnwidth]{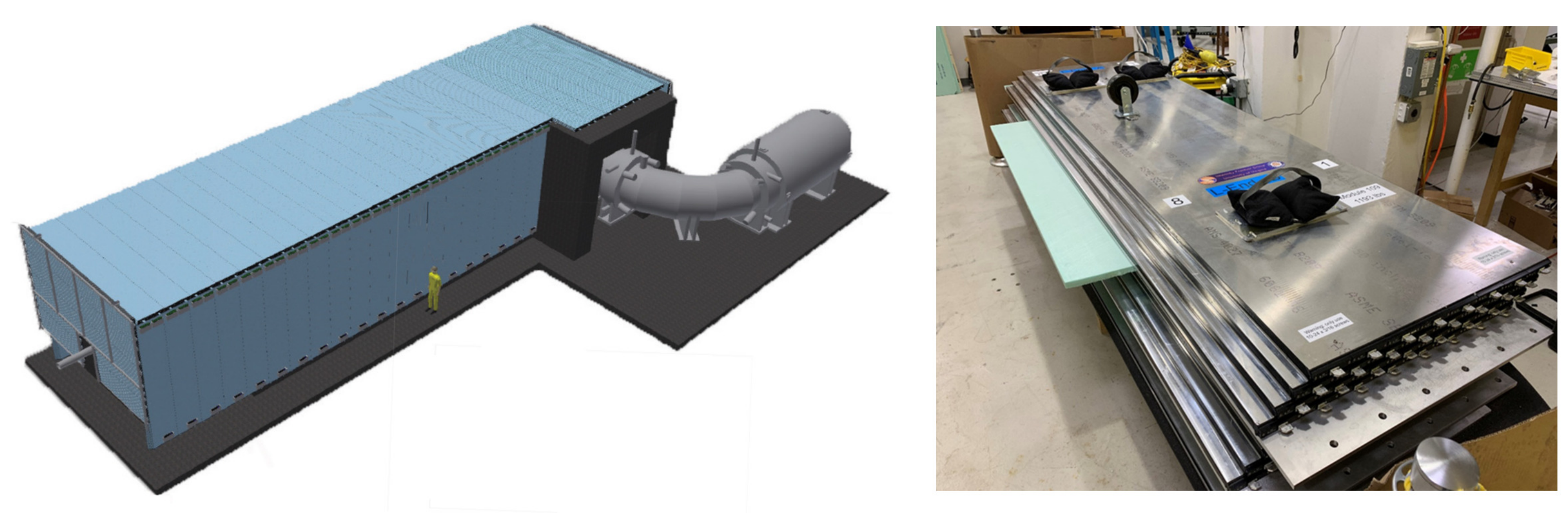}}
\vspace*{9pt}
\caption{Left: The CRV covers the DS and the downstream part of the TS. Right: Construction of a CRV module with four layers of scintillator counters.
\protect\label{fig:CRV_combined} }
\end{figure}

Backgrounds that mimic the CE signal can be generated by cosmic ray (CR) muons. 
The CR muon itself could mimic the signal in the tracker, if entering the DS at a shallow angle or scattering near the target. 
But particle identification at the calorimeter provides a muon rejection factor of 200, protecting against this class of backgrounds.
It is possible however that the CR muon will decay inside the DS or interact with material near the stopping target. 
Such processes can have available energy to generate a $\sim$\SI{105}{\MeV} electron, and are not prevented from occurring within the signal window.
It is estimated that one conversion-like electron per day can be produced by cosmic ray muons, which would yield $\mathcal{O}$(1,000) events over the expected run.
Therefore this background source must be suppressed by four orders of magnitude in order to achieve the sensitivity goals of Mu2e.

\begin{wrapfigure}{r}{0.57\textwidth}
	\centering
	\vspace*{8pt}
	\includegraphics[width=0.5\columnwidth]{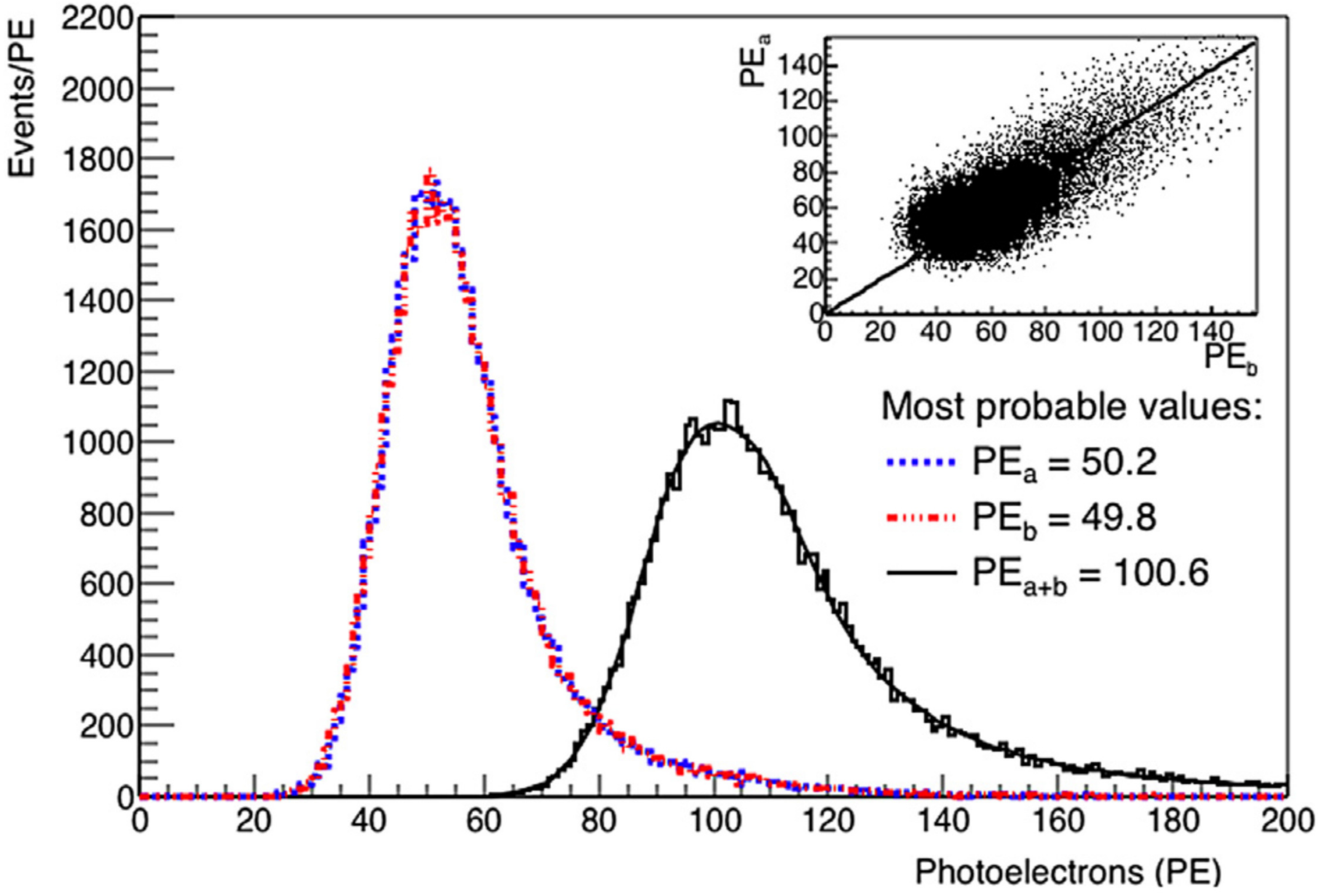}
	\caption{Photoelectron yield from the SiPM readout for a prototype counter. The responses from the SiPMs at one end of the counter (red and blue data) are shown to be properly correlated (insert). The summed response (black histogram) has a most probable value of $\sim$100 photoelectrons. From Artikov \etal\cite{Artikov2018}
	}
	\protect\label{fig:CRV_proto}
\end{wrapfigure}

Two mitigation strategies are adopted. 
The first is passive shielding, including the overburden above and to the sides of the detector hall, as well as the shielding concrete surrounding the Detector Solenoid.
The second and most important component is an active veto detector (cosmic ray veto - CRV) that will detect penetrating CR muons.

The CRV\cite{Group_2013} will encase the Detector Solenoid and a portion of the Transport Solenoid (Fig.~\ref{fig:CRV_combined}), covering a total area of about \SI{337}{\m\squared}.
The detecting element is extruded polystyrene scintillator counters coated with titanium dioxide (TiO$_2$), with embedded wavelength shifting fibers read out via SiPMs. 
The veto system employs four layers of these counters interspersed with aluminum absorbers. 
By requiring coincidence between at least three of the four layers the CRV can achieve 99.99\% efficiency, as required for adequate suppression of this background source.
Characterization of prototype counters in the Fermilab Test Beam Facility (Fig.~\ref{fig:CRV_proto}) determined the optimal concentration of TiO$_2$ coating and has demonstrated that the photoelectron yield meets specifications to achieve the required efficiency.\cite{Artikov2018}

\subsection{Other background sources}

\subsubsection{Antiproton-induced}

The \SI{8}{\GeV} primary proton beam is above the energy threshold for antiproton production through the $p + p \rightarrow p + p + \bar{p} + p$ process.
These antiprotons are a serious consideration as they do not decay, and some with low enough momentum can be transported through the solenoids.
Because their velocity is small, less than \SI{0.1}{\clight}, they travel slowly and can take microseconds to reach the stopping target.
They could then annihilate on material near the stopping target producing many secondary particles, which can generate a CE-mimicking $\sim$\SI{105}{\MeV} electron.
The most effective mitigation strategy is to block the antiprotons through a system of absorbers placed along the TS.
The absorbers are kept thin to maximize muon transmission with appropriate momentum selection.
The antiprotons have smaller kinetic energies and therefore suffer much larger $dE/dx$ losses, thus the system of absorbers accomplishes good suppression of antiprotons reaching the DS.

\begin{table}[b]
\tbl{Mean expectation for background events in Mu2e for \num{3.6e20} protons on target. When two numbers are quoted for uncertainty, the first corresponds to simulation statistics and the second to systematics. }
{\begin{tabular}{ccc}  
 \multicolumn{2}{c}{ \textbf{Background process} } 
& \textbf{Expected events} \\ 
\toprule
 Cosmic ray muons		&					&	$0.21 \pm 0.02 \pm 0.055$ \\
 \midrule
 \multirow{2}{*}{Intrinsic}	&	DIO				&	$0.14 \pm 0.03 \pm 0.11$ \\
 					&	RMC				&	$0.000_{-0.000}^{+0.004}$ \\
 \midrule
 \multirow{3}{*}{
 Prompt, late-arriving}	&	RPC				&	$0.021 \pm 0.001 \pm 0.002$ \\
 					&	Muon DIF			&	$<0.003$ \\
 					&	Pion DIF			&	$0.001 \pm <0.001$ \\
 					&	Beam electrons		&	$(2.1 \pm 1.0) \times 10^{-4}$ \\
 \midrule
 Antiproton-induced		&		 			&  $0.04 \pm 0.001 \pm 0.02$ \\ 
\botrule
\\
 \multicolumn{2}{c}{\textbf{Total}}						& \textbf{ 0.41 $\pm$ 0.03 (stat+syst) }\\ 
\end{tabular}
\protect\label{tab:Backgrounds} }
\end{table}

\subsubsection{Reconstruction errors}
\label{sec:recon_errors}

Track reconstruction can be affected by background activity in the tracker.
Such activity primarily originates from the muon beam, from multiple DIO electrons within a narrow time window, and from muon capture on a target nucleus that results in emission of photons, neutrons and protons. 
The ejected protons have a very small kinetic energy and are highly ionizing, inducing large pulses which should be identified by the straw ADC, but also increasing the dead time of hit straws. 
Delta ray emission is also possible following a proton hit, potentially triggering neighboring straws. 
Ejected neutrons can be captured on hydrogen or other atoms and produce low energy photons, begetting low momentum electrons which can generate a substantial number of in-time hits. 
This background activity scales linearly with beam intensity and causes tails in the resolution function that can push DIO electrons into the signal momentum window. 
The reconstruction software that must control for these resolution tails will be discussed in Section~\ref{sec:analysis}.

\subsection{Backgrounds summary and budget}

The expected total number of background events from each major source over the entire planned Mu2e operation is given in Table~\ref{tab:Backgrounds}.
These values have been estimated from Monte Carlo simulations over many Mu2e lifetimes, with cut selections aimed to optimize the experimental discovery sensitivity.
The total expected number of background events from all sources is $0.41 \pm 0.03$.
Thus any observation of an event consistent with a conversion electron would be an important suggestion of CLFV and new physics.
Using the techniques of Feldman and Cousins\cite{Feldman_1998} against the estimated background, the threshold for a $5 \sigma$ discovery would be the observation of more than $\sim$7.5 events.

\section{Simulation and Analysis}
\label{sec:analysis}

\begin{wrapfigure}{r}{0.67\textwidth}
	\centering
	\vspace*{8pt}
	\includegraphics[width=0.6\columnwidth]{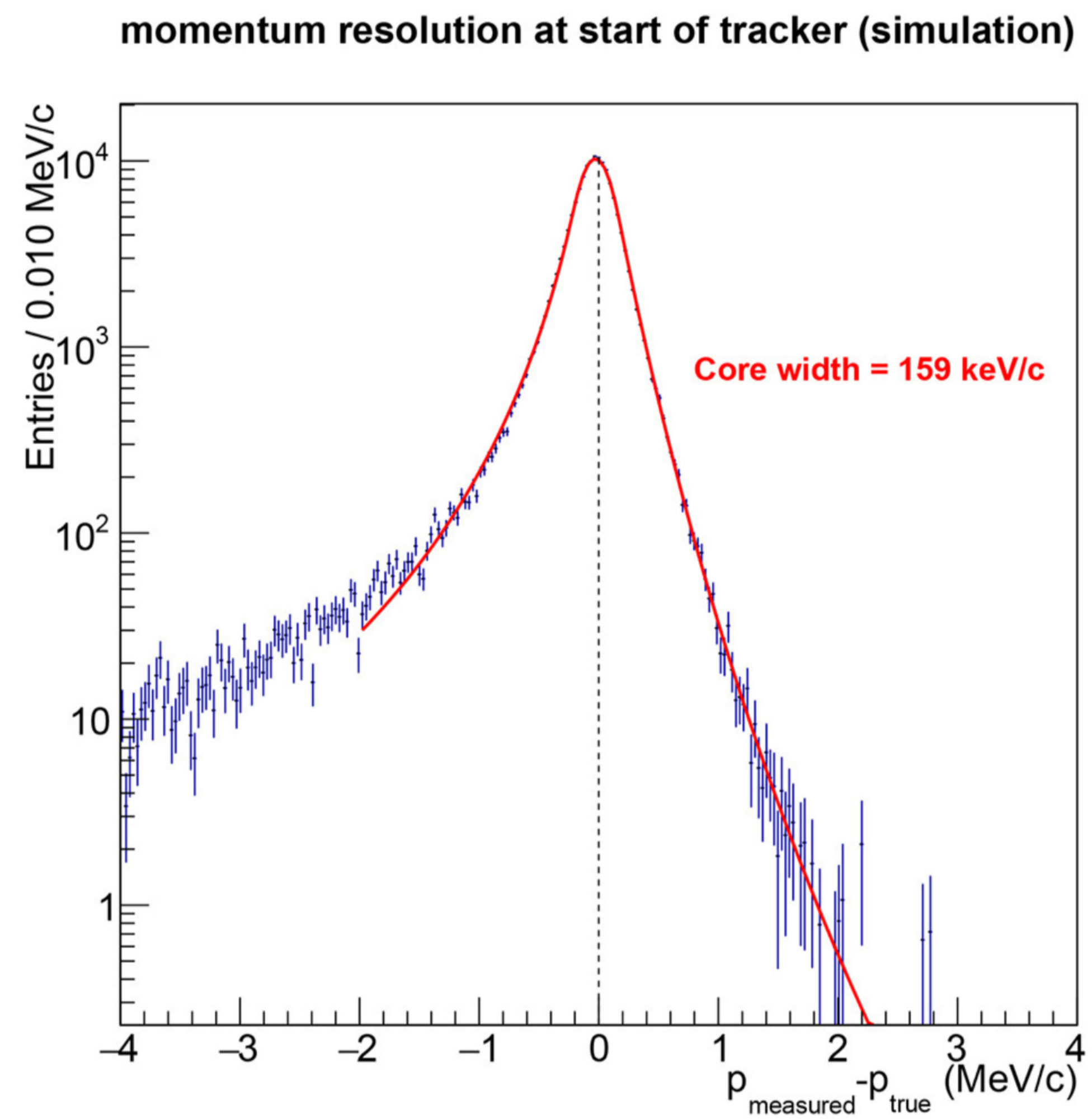}
	\caption{Resolution of the Mu2e tracker for electrons near the conversion energy, as the difference between the reconstructed momentum and the simulation truth. `Core width' refers to a fit in the central part of the resolution.
\protect\label{fig:tracker_resolution}}
\end{wrapfigure}

The Mu2e simulation software is based on the Geant4 package\cite{Geant4_2003} and aims to describe the geometry and materials on multiple scales, from the building walls and shielding to the multiple material layers of each tracker straw.
The decay in orbit is modeled after the Czarnecki spectrum\cite{Czarnecki_2011} and other processes are similarly based on published data and calculations.
Variations on these models are investigated as part of the systematic uncertainties quoted in Table~\ref{tab:Backgrounds}.
Since the tracker is the single most precise detector element and drives the experimental sensitivity a complete detailed simulation has been developed, from ionization in the straws to digitization output of the ADCs and TDCs on the tracker front-end.
The response has been tuned to fit the prototype test data of Fig.~\ref{fig:tracker_prototype_plots}.

Sophisticated and robust algorithms for pattern recognition and track fitting have been developed.
Cuts based on time, energy, and position of straw hits remove most of the proton and DIO background hits.
Time difference between hits from the same track is required to be within the maximum drift and transit time of \SI{50}{\ns}.
After significantly improving on signal/background ratio, straw hits are passed to a geometric pattern recognition algorithm. 
Initial helix parameters, covariance matrix, and track $t_0$ are produced to seed an iterative Kalman filter track fit.
The Kalman fit accounts for scattering and energy loss in the straw material, as well as inhomogeneity in the DS field, and returns the final reconstructed momentum evaluated at the upstream entrance to the tracker. 
The reconstructed trajectory must have an origin consistent with coming from the target, and the extrapolated track must have a calorimeter cluster that matches in time and position and is identified as an electron of consistent energy.

The simulated intrinsic momentum resolution of the tracker for signal conversion electrons that satisfy the track selection criteria, including material and reconstruction effects, is given in Fig.~\ref{fig:tracker_resolution}.
A core resolution of \SI{159}{\keVperc} is achieved. 
The high-side tail which could shift the fast-falling DIO spectrum to larger momentum is shown to be well controlled.
The momentum resolution is found to be well within experimental requirements, and also robust against simulated increases in rate.
It remains a subdominant contribution to energy loss and resolution smearing due to interactions in material upstream of the tracker.

\begin{figure}[t]
\centerline{\includegraphics[width=0.85\columnwidth]{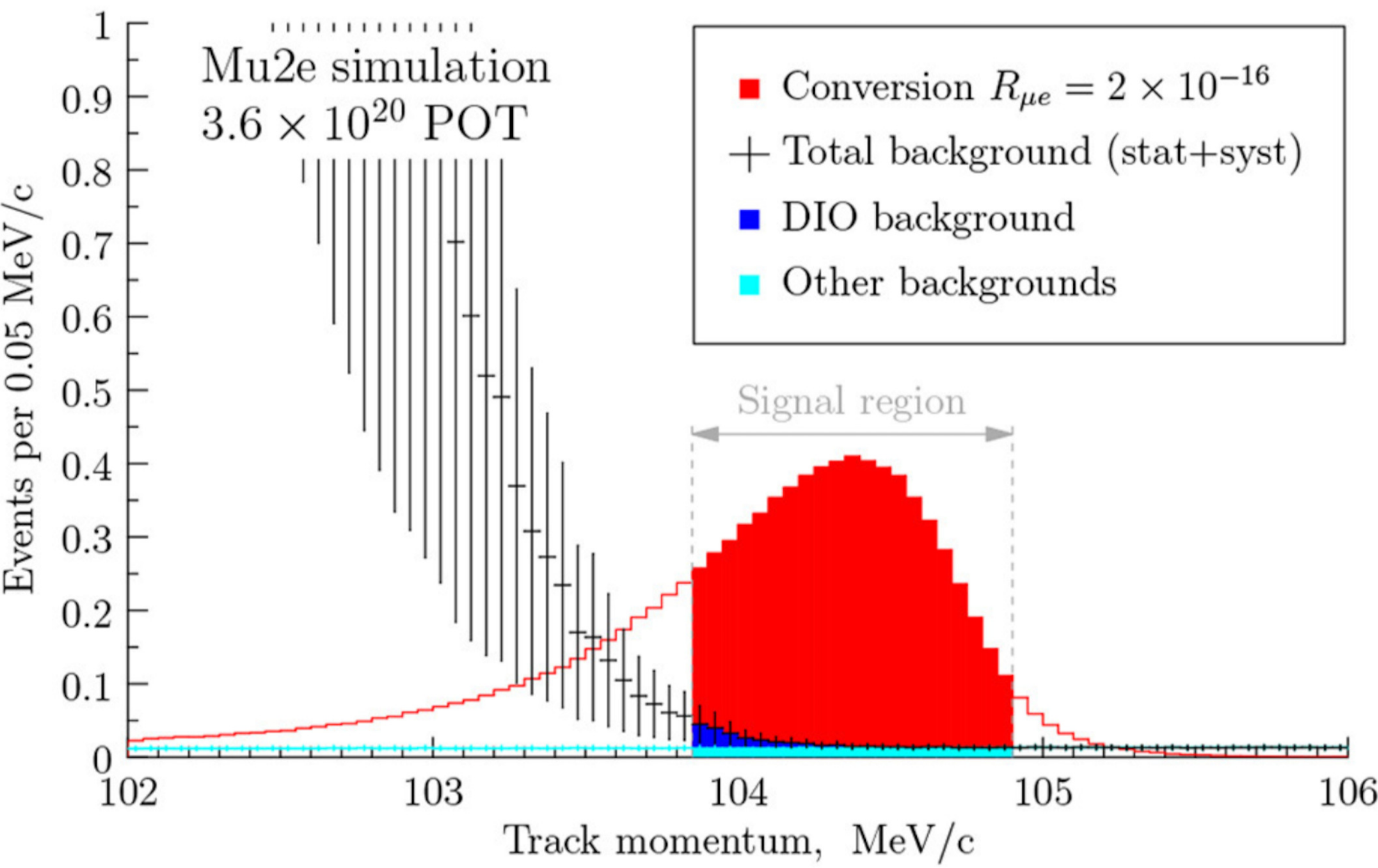}}
\vspace*{9pt}
\caption{Conversion signal and DIO background probability distributions for a simulated Mu2e experiment assuming $R_{\mu e} =$ \num{2e-16}.  \protect\label{fig:signal_pseudoexperiment}}
\end{figure}

Fig.~\ref{fig:signal_pseudoexperiment} presents the momentum probability distributions for the DIO spectrum and the conversion signal, using realistic acceptance and resolution in a pseudo-experiment with $R_{\mu e} =$ \num{2e-16}.
The distributions of both DIO and the monochromatic CE signal are shifted and smeared due to interactions in material and detector resolution, resulting in partial overlap.
An optimal signal region of [103.85, 104.90] \SI{}{\MeVperc} is defined to optimize discovery sensitivity by avoiding most of the region of overlap. Only an acceptably small DIO component is allowed within the momentum window, which integrates to the expectation of $0.14 \pm 0.03\text{(stat)} \pm 0.11\text{(syst)}$ DIO events (from Table~\ref{tab:Backgrounds}) over the lifetime of the experiment.
The upper bound of the selection region limits contributions from other background sources which scale with window size, like those from cosmic ray muons.

Using the techniques of Feldman and Cousins\cite{Feldman_1998} against the estimated background of 0.41 events, a $5 \sigma$ discovery could be claimed for 7.5 events within the signal window. 
In fact this is exactly the number of events expected in the pseudo-experiment of Fig.~\ref{fig:signal_pseudoexperiment}; that defines the Mu2e $5 \sigma$ discovery sensitivity at $R_{\mu e} =$ \SI{2e-16}, orders of magnitude beyond current constraints.
Mu2e will have a single-event sensitivity, \ie an expectation to observe a single conversion event within the signal window, for $R_{\mu e} = \num{3e-17} $.

\section{Status and Outlook}

The Mu2e project is currently actively in the construction phase.
For the solenoids, which are the main schedule driver, the entire length of superconducting cable has been procured and tested. Winding is in progress for all three solenoid units.
The heat and radiation shield has been constructed and delivered. 
All 30,000 (including spares) tracker straws have been procured and panel manufacturing is in progress. 
All tracker FEE boards have been prototyped and tested and are ready for production.
Nearly all CsI crystals for the calorimeter have been delivered and tested.
Extrusion fabrication is completed for the CRV counters, which are being assembled into layers.

\begin{figure}[b]
\centerline{\includegraphics[width=\columnwidth]{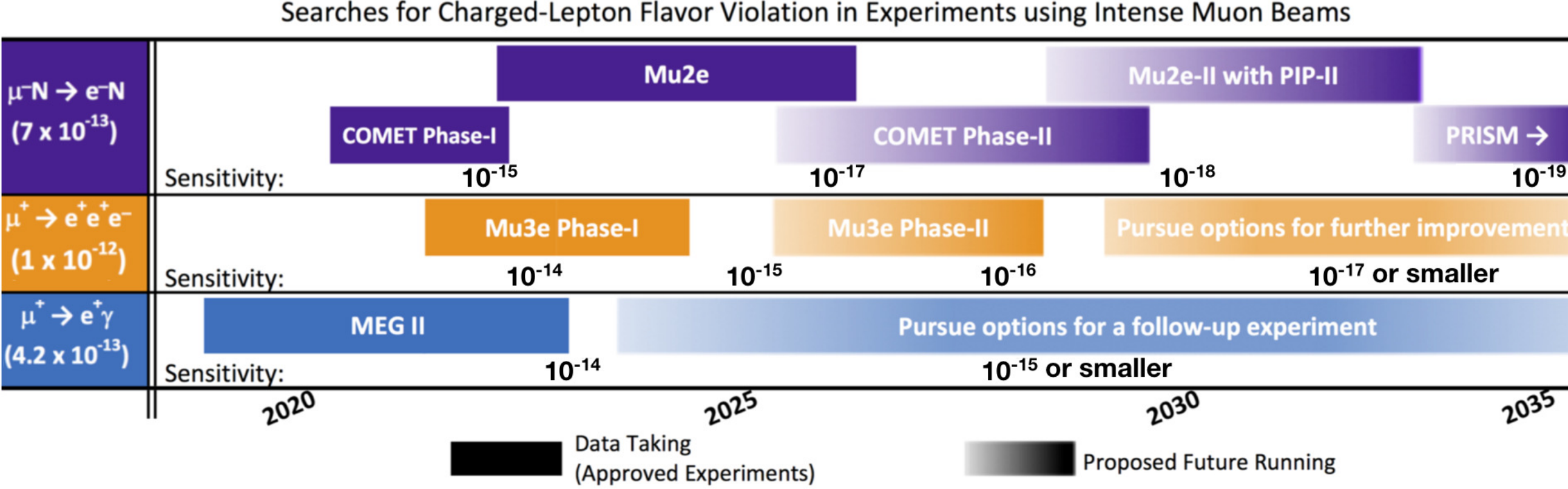}}
\vspace*{9pt}
\caption{
	Planned schedules for current searches for charged lepton flavor violating $\mu \rightarrow e$ transitions. 
	Also shown are possible schedules for proposed upgrades to these experiments. 
	The current best limits for each process are shown on the left in parentheses, while expected future sensitivities are indicated by order of magnitude along the bottom of each row. From Baldini \etal\cite{Baldini_2019elc}
\protect\label{fig:timeline}}
\end{figure}

The Mu2e project will transition to installation in 2020-2021, with commissioning between 2021-2022. 
Physics running is expected to begin in 2023, followed by three years of operations.
By that time the upgraded MEG measurement at PSI,\cite{MEG_2018} as well as Phase-I of the Mu3e\cite{Mu3e_2013} and COMET\cite{COMET_2009} experiments, are expected to have first results. 
Mu2e is expected to further increase sensitivity by another two orders of magnitude beyond these measurements, culminating an intense program over the next years that may well yield the first discovery of CLFV events. The planned timeline of this experimental program is shown in Figure~\ref{fig:timeline}.
If a signal is observed then the complementarity between the three muon transition modes will be a powerful discriminant among the underlying New Physics models.

\subsubsection*{Mu2e-II}

Looking further into the future, another order of magnitude improvement in sensitivity beyond Mu2e is possible as the conversion channel is unburdened by accidental coincidences. 
In Fermilab the second phase of the Proton Improvement Plan (PIP-II) would provide unique capabilities and could deliver \SI{80}{kW} beam power to an upgraded Mu2e-II search with over \num{e11} stopped $\mu^{-}/s$.\cite{Mu2eII_2018}
The delivered beam would be near \SI{1}{\GeV} in energy to eliminate backgrounds from antiproton production.
The internal DIO background would scale with beam intensity, therefore improved resolution would be required to maintain  control of this contribution. 
If Mu2e does not observe a signal then an increase in sensitivity by another order of magnitude would be a powerful search over a wide area of renewed phase space. 
If a CLFV signal is indeed observed, in a discovery of new physics that parallels that of neutrino oscillations, then the Mu2e-II upgrade would be necessary to increase the statistical significance of the conversion rate. 
New possibilities open up to measure the conversion rate on different stopping nuclei, gaining discriminating power between models of new physics that could generate the conversion process.\cite{Cirigliano_2009}
The proposed Mu2e-II search would have the highest sensitivity to new physics inducing charged lepton flavor violation for the foreseeable future.

\section*{Acknowledgments}

This manuscript has been authored by Fermi Research Alliance, LLC under Contract No. DE-AC02-07CH11359 with the U.S. Department of Energy, Office of Science, Office of High Energy Physics.
The author would like to acknowledge the generous support by the Mu2e collaboration, especially that of Doug Glenzinski and Robert Bernstein to early-career collaborators.
Many thanks also to Simona Giovanella for her valuable feedback on this manuscript.

\bibliographystyle{unsrt}
\bibliography{Mu2e_paper}

\end{document}